\documentclass{CRPITStyle} 

\usepackage[authoryear]{natbib}
\renewcommand{\cite}{\citep}
\usepackage{xspace}
\usepackage{epsfig}

\usepackage{amsmath, amsthm, amssymb}

\begin{document}

\title{Comparing RNA structures using a full set of \\
  biologically relevant edit operations is intractable}
\author{Guillaume Blin$^1$
\and 
Sylvie Hamel$^2$
\and 
St\'ephane Vialette$^1$}
\affiliation{$^1$ Universit\'e Paris-Est, IGM-LabInfo - UMR CNRS 8049, France\\
Email:~{\tt \{gblin,vialette\}@univ-mlv.fr}\\[.1in]
$^2$ DIRO - Universit\'e de Montr\'eal - QC - Canada\\
Email:~{\tt hamelsyl@iro.umontreal.ca}}

\maketitle

\newcommand \edit {\mbox{\PB{Edit}}\xspace}
\newcommand \lapcs {\mbox{\sc{Lapcs}}}
\newcommand \longlapcs {\sc{Longest} \sc{Arc-Preserving} \sc{Common} \sc{Subsequence}}
\newcommand \stem {\mbox{\sc{Stem}}\xspace}
\newcommand \nested {\mbox{\sc{Nested}}\xspace}
\newcommand \plain {\mbox{\sc{Plain}}\xspace}
\newcommand \crossing {\mbox{\sc{Crossing}}\xspace}
\newcommand \unlimited {\mbox{\sc{Unlimited}}\xspace}
\newcommand \unlim {\mbox{\sc{Unlim}}\xspace}
\newcommand \cros {\mbox{\sc{Cros}}\xspace}
\newcommand \nest {\mbox{\sc{Nest}}\xspace}

\newtheorem{theorem}{Theorem}
\newtheorem{corollary}{Corollary}
\newtheorem{lemma}{Lemma}



\newcommand\conferencenameandplace{15th Computing: The Australasian Theory Symposium (CATS2009), Wellington, New Zealand}
\newcommand\volumenumber{XX}
\newcommand\conferenceyear{2008}
\newcommand\editorname{Rod Downey and Prabhu Manyem}
\toappearstandard 


\begin{abstract}
Arc-annotated sequences are useful for representing structural
information of RNAs and have been extensively used for 
comparing RNA structures in both terms of sequence and structural
similarities. Among the many paradigms referring to arc-annotated
sequences and RNA structures comparison (see \cite{IGMA_BliDenDul08}
for more details), the most important one is the general edit
distance. 
The problem of computing an edit distance between two
non-crossing arc-annotated sequences was introduced in \cite{Evans99}.
The introduced model uses edit operations that involve 
either single letters or pairs
of letters (never considered separately)
and is solvable in
polynomial-time \cite{ZhangShasha:1989}.  

To account for other possible RNA structural evolutionary events, new
edit operations, allowing to consider either silmutaneously or
separately letters of a pair were introduced in \cite{jiangli};
unfortunately at the cost of computational tractability. It has been
proved that comparing two RNA secondary structures using a full set of
biologically relevant edit operations is {\sf\bf
  NP}-complete. Nevertheless, in \cite{DBLP:conf/spire/GuignonCH05},
the authors have used a strong combinatorial restriction in order to
compare two RNA stem-loops with a full set of biologically relevant
edit operations; which have allowed them to design 
a polynomial-time and space algorithm for comparing general secondary
RNA structures. 

In this paper we will prove theoretically that comparing two RNA
structures using a full set of biologically relevant edit operations
cannot be done without strong combinatorial restrictions. 
\end{abstract}
\vspace{.1in}


\section{Introduction}
In computational biology, comparison of RNA molecules has recently
attracted a lot of interest  due to the  rapidly increasing amount of
known RNA molecules, especially non-coding RNAs. Very often, 
\emph{arc-annotated sequences}, originally introduced in
\cite{Evans99}, 
are used to represent RNA structures. An arc-annotated sequence is a
sequence over a given alphabet together with additional structural
information specified by arcs connecting pairs
of positions. 
The arcs determine the  way the sequence folds into a
three-dimensional space.  

The problem of computing an edit distance between two arc-annotated
sequences was introduced in  \cite{Evans99} with a model that used
only three edit operations (deletion, insertion and substitution) 
either on single letters (letters in the sequence with no
incident arc) or pairs of letters (letters connected by an
arc). 
In this model, the two letters of an arc are
never considered separately, and hence the problem of computing the
edit distance between two arc-annotated sequences becomes equivalent
(when no pair of 
arcs are crossing) to the tree edit distance problem, that can be
solved in polynomial-time \cite{ZhangShasha:1989}.  

To account for other possible RNA structural evolutionary events, new
edit operations, such as creation, deletion or modification of arcs
between pairs of letters, were introduced in \cite{jiangli} at the 
cost of computational tractability. Indeed, it has been shown in
\cite{IGMA_BliFerRus07} that in case of non-crossing arcs, the problem
of computing the edit distance between two 
arc-annotated sequences under this model is {\sf\bf NP}-hard. 
Playing the game of applying constraints either
on the legal edit operations  
or on the allowed alignments, 
several papers have shed new light on 
the borderline between tractability and intractability
\cite{DBLP:conf/spire/GuignonCH05,IGMA_BliDenDul08}. 
Of particular importance, in \cite{DBLP:conf/spire/GuignonCH05}, the
authors introduced the notion  
of \textit{conservative edit distance and mapping} between two RNA
stem-loops in order to design a polynomial-time algorithm
for comparing general secondary RNA structures using the full set of
biological edit operations introduced in \cite{jiangli}. 
This algorithm  
is based on a decomposition in stem-loop-like substructures that are
pairwised compared and  
used to compare complete RNA secondary structures. As mentionned in
\cite{DBLP:conf/spire/GuignonCH05},  
whereas in the very restrictive case of conservative distance and mapping, the 
computation of the general edit distance is polynomial-time solvable,
it is not known if the general, \emph{i.e.}, not conservative,
edit distance between two stem-loops can be also computed in
polynomial-time.  

In this paper, we will show that this strong combinatorial restriction
was necessary for the problem 
to become polynomial since it is {\sf\bf NP}-hard in the general
case. Despite the fact that this result may 
be considered as purely theoretical, 
it proves that comparing two  
RNA structures using a full set of biologically relevant edit
operations cannot be done without strong  
combinatorial restrictions.

\section{Preliminaries}
Given a finite alphabet $\Sigma$, an arc-annotated sequence is
formally defined 
by a pair $(S, P)$, where $S$ is a string of  $\Sigma^*$ and $P$ is a
set of arcs connecting pairs of letters of $S$. In reference to RNA
structures, letters are called \emph{bases}. Bases with no incident
arc are called \emph{single bases}. In an arc-annotated sequence, two
arcs $(i_1 , j_1)$ and $(i_2 , j_2)$ are crossing, if  
$i_1 < i_2 < j_1 < j_2$ or $i_2 < i_1 < j_2 < j_1$.  
An arc $(i_1, j_1)$ is \emph{embedded} into
another arc $(i_2, j_2)$ if
$i_2 < i_1 < j_1 < j_2$.
Evans \cite{Evans99} 
(see \cite{DBLP:conf/spire/GuignonCH05} for extensions)
introduced five different levels of arc structure:
{\unlimited} -- no restriction at all;  
{\crossing} -- there is no base incident to more than one arc; 
{\nested}  -- there is no base incident to more than one arc and no
two arcs are crossing;
{\stem} -- there is no base incident to more than one arc and given
any two arcs, one is embedded into the other;   
{\plain} -- there is no arc. There is an obvious inclusion relation
between those levels: 
$\plain \subset \stem \subset {\nested} \subset \crossing \subset
\unlimited$. 
An arc-annotated sequence $(S_1, P_1)$ is said to \emph{occur} in another
arc-annotated sequence $(S_2, P_2)$ if one can obtain the former from
the latter by repeatedly deleting bases (deleting a base that is
incident to an arc results in the deletion of the arc).

Among the many paradigms referring to arc-annotated sequences (see
\cite{IGMA_BliDenDul08} for more details) we focus in this article on
the {\longlapcs} ({\lapcs} for short) \cite{Evans99,jiangma,lin2002}
and the general edit distance ({\textsc{Edit}} for short)  
\cite{jiangli,IGMA_BliFerHer07}. Indeed, as shown in
\cite{IGMA_BliDenDul08}, those two 
paradigms are quite related since the {\lapcs} problem is a special
case of {\textsc{Edit}} when  
considering the complete set of edit operations defined in
\cite{jiangli}. Therefore, the  
hardness results for {\lapcs} stands for {\textsc{Edit}}. 

Formally,  the {\longlapcs}
problem  is defined as follows: given two
arc-annotated sequences  $(S_1,P_1)$ and  $(S_2,P_2)$, find the
longest -- in terms of sequence length -- common arc-annotated
subsequence that occurs in both $(S_1,P_1)$ and
$(S_2,P_2)$.
It has been shown in \cite{jiangli}
that the {\lapcs} problem is {{\sf\bf NP}-hard} even for {\nested}
structures, \emph{i.e.},
\lapcs(\nested, \nested). 
Still focussing on \nested structures, 
Alber \emph{et al.}
\cite{Alber:Gramm:Guo:Niedermeier:TCS:2004} proved 
that the \lapcs(\nested, \nested) problem is solvable in  
$O(3^k \, |\Sigma|^k \, kn)$ time, 
where $n$ is the maximum length of the two
sequences and $k$ is the length of the common subsequence searched
for.    
The $O(3^k \, |\Sigma|^k \, kn)$ time parameterized algorithm by 
Alber \emph{et al.} is by brute-force enumeration:
(i)
Generate all possible sequences of length $k$ with all
possible $\nested$ arc annotations, and
(ii)
For each of these arc-annotated candidate sequences, 
check whether or not it occurs as a pattern in both $S_1$ and
$S_2$.
At the heart of this approach is the fact that 
it can be decided in $O(n \, k)$ time 
whether or not this sequence occurs as an arc-preserving common subsequence
\cite{Gramm:Guo:Niedermeier:ACM-TransAlgo:2006}. 
It is easily see that the above algorithm reduces to
$O(2^{3k-1} \, km)$ time for \lapcs(\stem, \stem).
Indeed, there exist $|\Sigma|^k$ sequences of length $k$
and hence, for a given sequence of length $k$, 
there exist $\binom{k}{i}$ different arc-annotations 
with $i$ arcs.
Therefore, there exist
$
\sum_{i=0}^{\left\lfloor k/2 \right\rfloor} \binom{k}{2i} 
=
2^{k-1}
$
arc-annotations of a given sequence of length $k$.

Here, we focus on the only remaining open problems concerning
{\lapcs}  and  {\textsc{Edit}} over stem-loops by 
showing, with a unique proof, their hardness. More precisely, we prove
that \lapcs(\stem, \stem) - which 
may be considered as a very restricted problem and thus not
interesting - is {\sf\bf NP}-hard \textbf{in order to infer} 
the {\sf\bf NP}-hardness of \textsc{Edit}(\stem, \stem) - which is for
sure, according to \cite{DBLP:conf/spire/GuignonCH05},  
an interesting problem that can be used in a very simple way to
compare complete RNA secondary structures.  
This results also prove that in any future work on comparing RNA
structures with a full set of edit operations  
it will be necessary to introduce strong combinatorial restrictions in
order to get an exact polynomial-time algorithm  
since even with the simpliest model, the general edit distance problem is still {\sf\bf NP}-complete.

\section{Comparing RNA Stem-Loops is {\sf\bf NP}-complete}\label{SectionNPC}

In this section, we prove that {\lapcs} over stem-loops
($\lapcs(\stem,\stem)$) is {\sf\bf NP}-complete (in Theorem
\ref{thm_lapcs}); therefore answering an open question of
\cite{DBLP:conf/spire/GuignonCH05}. This last result induces the
{\sf\bf NP}-hardness of $\textsc{Edit}$ over stem-loops. 

\begin{theorem}\label{thm_lapcs}
 $\lapcs(\stem,\stem)$ is {\sf\bf NP}-complete.
\end{theorem}

\begin{corollary}
 Comparing RNA structures with a full set of biologically relevant
 edit operations cannot be done without introducing strong
 combinatorial restrictions. 
\end{corollary}

 In the following, we consider the decision version of the problem which corresponds to deciding if there exists an arc-preserving common subsequence of length greater or equal to a given parameter $k'$.

It is easy to see that the $\lapcs$ problem is in ${\sf\bf NP}$. In order
to prove its ${\sf\bf NP}-hardness$, we define a reduction from the {\sf\bf NP}-complete 3SAT
problem \cite{Garey:Johnson:1979} which is defined as follows: Given a
collection $C_q = \{c_1,$ $c_2,$ $\ldots ,$ $c_q\}$ of $q$ clauses,
where each clause consists of a set of 3 literals (representing the
disjunction of those literals) over a finite set of $n$ boolean
variables $V_n = \{x_1,$ $x_2,$ $\ldots ,$ $x_n\}$, is there an
assignment of truth values to each variable of $V_n$ s.t. at least one
of the literals in each clause is true?  

Let $(C_q,V_n)$ be any instance of the 3SAT problem s.t. $C_q =
\{c_1,$ $c_2,$ $\ldots ,$ $c_q\}$ and $V_n = \{x_1,$ $x_2,$ $\ldots ,$
$x_n\}$. For convenience, let $L^j_i$ denote the $j^{th}$ literal of
the $i^{th}$ clause (i.e. $c_i$) of $C_q$. In the following, given a
sequence $S$ over an alphabet $\Sigma$, let $\chi(i,c,S)$ denote the
$i^{th}$ occurrence of the letter $c$ in $S$.  

\begin{figure*}
\begin{center}
 \includegraphics[angle=90,height=20cm]{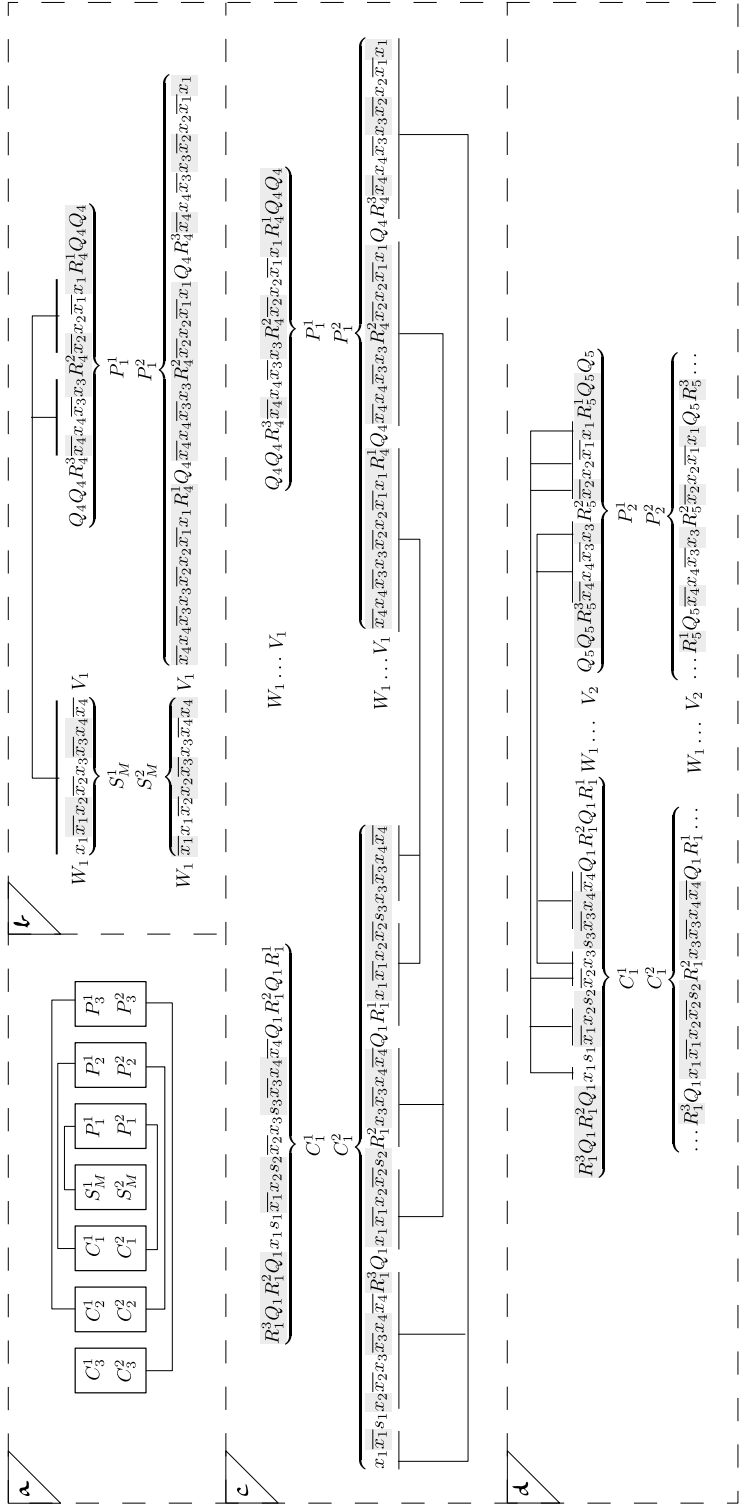}
\caption{Considering $C_q=(x_1\vee x_2\vee \overline{x_3})\wedge(\overline{x_1}\vee \overline{x_2}\vee x_4)\wedge(x_2\vee \overline{x_3}\vee \overline{x_4})$. For readability, all the arcs have not been drawn, consecutive arcs are representing by a unique arc with lines for endpoints. Symbols over a grey background may be deleted to obtain an optimal LAPCS. a) A schematic view of the overall arrangement of the components of the two a.a. sequences. b) Description of $S_M^1$, $S_M^2$, $P_1^1$, $P_1^2$ and the corresponding arcs in $P_1$. c) Description of $C_1^1$, $C_1^2$, $P_1^1$, $P_1^2$ and the corresponding arcs in $P_2$. d) Description of $C_1^1$, $C_1^2$, $P_2^1$, $P_2^2$ and the corresponding arcs in $P_1$.}\label{Fig_Example_1}
\end{center}
\end{figure*}

\begin{figure*}
\begin{center}
 \includegraphics[angle=90,height=20cm]{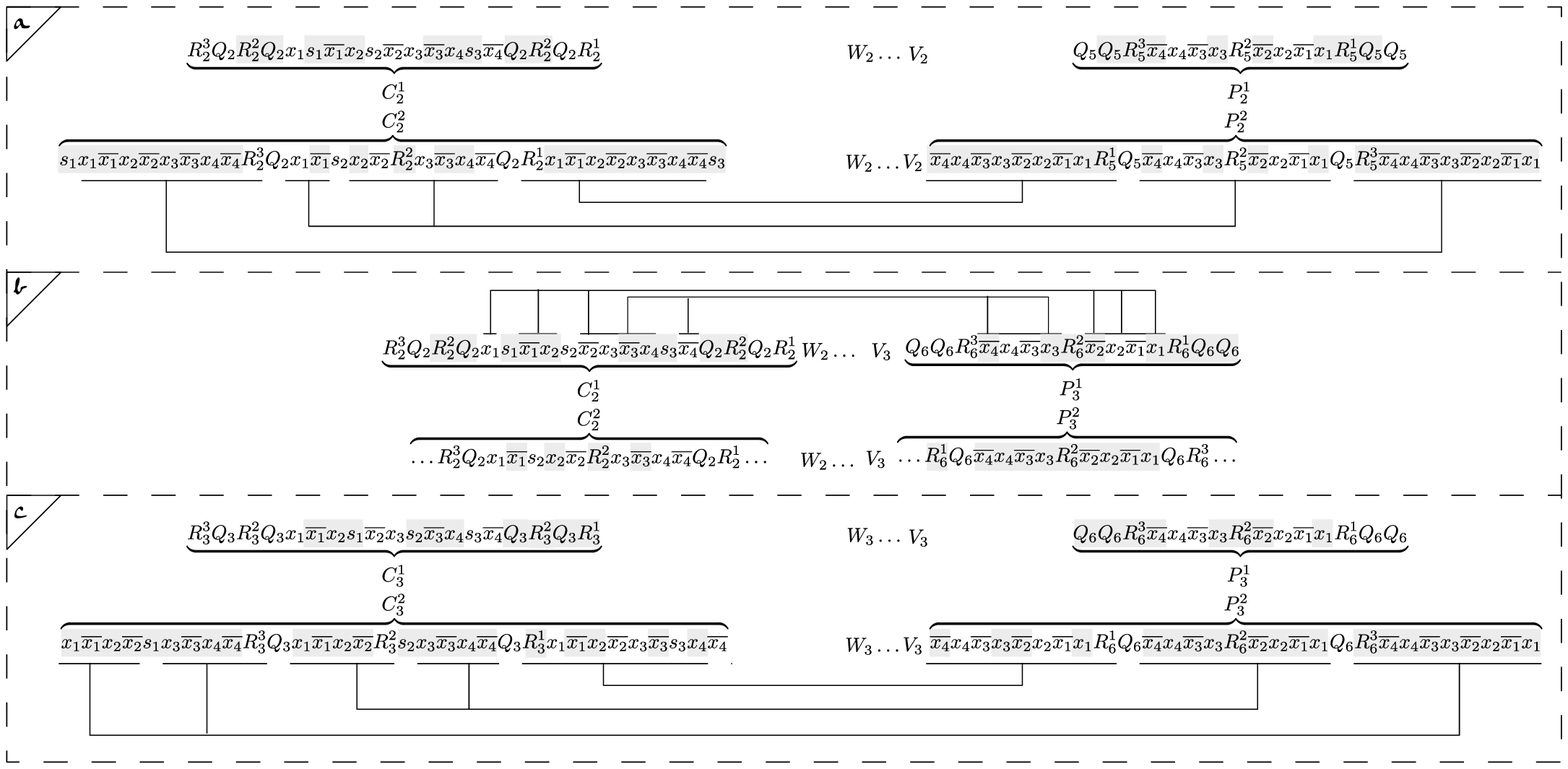}
\caption{Considering $C_q=(x_1\vee x_2\vee \overline{x_3})\wedge(\overline{x_1}\vee \overline{x_2}\vee x_4)\wedge(x_2\vee \overline{x_3}\vee \overline{x_4})$. For readability all the arcs have not been drawn, consecutive arcs are representing by a unique arc with lines for endpoints. Symbols over a grey background may be deleted to obtain an optimal LAPCS. a) Description of $C_2^1$, $C_2^2$, $P_2^1$, $P_2^2$ and the corresponding arcs in $P_2$. c) Description of $C_2^1$, $C_2^2$, $P_3^1$, $P_3^2$ and the corresponding arcs in $P_1$. d) Description of $C_3^1$, $C_3^2$, $P_3^1$, $P_3^2$ and the corresponding arcs in $P_2$.}\label{Fig_Example_2}
\end{center}
\end{figure*}

We build two arc-annotated sequences $(S_1,P_1)$ and  $(S_2,P_2)$ as
follows. An illustration of a full example is given in Figures
\ref{Fig_Example_1} and \ref{Fig_Example_2}, where $n=4$ and
$q=3$. For readability reasons, the arc-annotated sequences resulting
from the construction have been split into several parts and a
schematic overview of the overall placement of each part is provided.

Let $S_1=C_q^1W_qC_{q-1}^1\ldots C_2^1W_2C_1^1W_1S_M^1V_1P_1^1V_2$ $P_2^1\ldots P_{q-1}^1V_qP_q^1$ and $S_2=C_q^2W_qC_{q-1}^2\ldots C_2^2W_2C_1^2W_1$ $S_M^2$ $V_1P_1^2V_2P_2^2\ldots P_{q-1}^2V_qP_q^2$ such that for all $1\leq i \leq q, 1\leq k \leq n$, 
	\begin{itemize}
        \item $C_i^1 = R_i^3Q_iR^2_iQ_iX^1_1X^1_2\ldots$ $X^1_nQ_iR_i^2Q_iR^1_i$ with $X^1_k=x_ks_j\overline{x_k}$ if $x_k=L^j_i$ or $\overline{x_k}=L^j_i$; $X^1_k=x_k\overline{x_k}$ otherwise;
        \item $P_i^1 =Q_{q+i}Q_{q+i}R_{q+i}^3X^1_n\ldots X^1_{\frac{n}{2}+1} R_{q+i}^2X^1_{\frac{n}{2}}\ldots X^1_1$ $R_{q+i}^1Q_{q+i}Q_{q+i}$ such that $X^1_k=\overline{x_k}x_k$;
        \item $C_i^2 = X^2_1\ldots X^2_nR_i^3Q_iX^2_1\ldots$ $X^2_{\frac{n}{2}}R^2_iX^2_{\frac{n}{2}+1}\ldots X^2_1$ $Q_iR_i^1X^2_1\ldots X^2_n$ such that for $1\leq j \leq 3$, $\chi(j,X^2_k,C_i^2)=x_k\overline{x_k}s_j$ (resp. $s_jx_k\overline{x_k}$) if $x_k=L^j_i$ (resp. $\overline{x_k}=L^j_i$); $\chi(j,X^2_k,C_i^2)=x_k\overline{x_k}$ otherwise;
        \item $P_i^2 =X^2_n\ldots X^2_1R_{q+i}^1Q_{q+i}X^2_n\ldots X^2_{\frac{n}{2}+1} R_{q+i}^2$ $X^2_{\frac{n}{2}}\ldots X^2_1Q_{q+i}R_{q+i}^3X^2_n\ldots X^2_1$ with $X^2_k=\overline{x_k}x_k$.
	\end{itemize}
Moreover, let $S_M^1=x_1\overline{x_1}x_2\overline{x_2}\ldots x_n\overline{x_n}$ and $S_M^2=\overline{x_1}x_1\overline{x_2}x_2\ldots \overline{x_n}x_n$. Notice that, by construction, there is only one occurrence of each $\{s_1,s_2,s_3\}$ in $C^2_i$.

For all $1\leq i \leq q$, let $Q_i$ (resp. $Q_{q+i}$) be a segment of $n+1$ symbols $y_i$ (resp. $y_{q+i}$). Moreover, for all $1 \leq i \leq q$, let $W_i$ (resp. $V_i$) be a segment of $20(max\{q,n\}^2)$ symbols $w_i$ (resp. $v_i$). Let us now define $P_1$ and $P_2$. 

For all $1\leq i \leq q-1$, (1) add an arc in $P_1$ between $\chi(1,x_k,C_i^1)$ (resp. $\chi(1,\overline{x_k},C_i^1)$) and $\chi(1,x_k,P_{i+1}^1)$ (resp. $\chi(1,\overline{x_k},P_{i+1}^1)$), $\forall 1\leq k \leq n$ (see Figure \ref{Fig_Example_1}.d and \ref{Fig_Example_2}.b); (2) add an arc in $P_2$ between $\chi(j,x_k,C_i^2)$ (resp. $\chi(j,\overline{x_k},C_i^2)$) and $\chi((4-j),x_k,P_i^2)$ (resp. $\chi((4-j),\overline{x_k},P_i^2)$), $\forall 1\leq k \leq n$ (see Figure \ref{Fig_Example_1}.c, \ref{Fig_Example_2}.a and \ref{Fig_Example_2}.c); (3) add an arc in $P_2$ between $\chi(1,R_i^j,C_i^2)$ and $\chi(1,R_{q+i}^j,P_i^2)$, $\forall 1\leq j \leq 3$ (see Figure \ref{Fig_Example_1}.c, \ref{Fig_Example_2}.a and \ref{Fig_Example_2}.c).

Clearly, this construction can be achieved in polynomial-time, and yields to sequences $(S_1,P_1)$ and $(S_2,P_2)$ that are both of type {\stem}. We now give an intuitive description of the different elements of this construction.

Each clause $c_i\in C_q$ is represented by a pair $(C^1_i,C^2_i)$ of sequences. The sequence $C^2_i$ is composed of 
three subsequences representing a selection mechanism of one of the three literals of $c_i$. The pair $(S_M^1,S_M^2)$ of sequences is a control mechanism that will guarantee that a variable $x_k$ cannot be true and false simultaneously. Finally, for each clause $c_i\in C_q$, the pair $(P^1_i,P^2_i)$ of sequences is a propagation mechanism which aim is to propagate the selection of the assignment (i.e. true or false) of any literal $x_k$ all over $C_q$. Notice that all the previous intuitive notions will be detailed and clarified afterwards. 

In the rest of this article, we will refer to any such construction as a \textit{snail-construction}. In order to complete the instance of the $\lapcs(\stem,\stem)$ problem, we define the parameter $k' = 40q(max\{q,n\}^2)+6qn+8q+n$ which corresponds to the desired length of the solution. In the following, let $(S_1,P_1)$ and $(S_2,P_2)$ denote the arc-annotated sequences obtained by a snail-construction. We will denote $S_d$ the set of symbols deleted in a solution of $\lapcs$ problem on $(S_1,P_1)$ and $(S_2,P_2)$ (i.e. the symbols that do not belong to the common subsequence).

We start the proof that the reduction from 3SAT to $\lapcs(\stem,\stem)$ is correct by giving some properties about any optimal solution.

\begin{lemma}\label{lem_canonicity}
 In any optimal solution of $\lapcs$ problem on $(S_1,P_1)$ and $(S_2,P_2)$, at least one symbol incident to any arc would be deleted. Moreover, all the symbols of $V_i$ and $W_i$, for $1\leq i \leq q$, will not be deleted.
\end{lemma}
\begin{proof}
 By contradiction, let us suppose that there exist at least one arc s.t. the two symbols incident to this last are not deleted in a solution of $\lapcs$ problem on $(S_1,P_1)$ and $(S_2,P_2)$. Then, by construction, it induces that at least one complete sequence $V_j$ or $W_j$, for a given $1\leq j \leq q$, has been deleted. Since they have the same length, we will consider w.l.o.g. afterwards that $V_i$ has been deleted. Therefore, since $S_1$ is, by construction, smaller than $S_2$ the length of this optimal solution is at most $|S_1| - |V_j| = \sum_{i=1}^{q}(|C^1_i|+|P^1_i|+|V_i|+|W_i|) + |S^1_M| - |V_j| = \sum_{i=1}^{q}((6n+11)+(6n+7)+(20(max\{q,n\}^2))+(20(max\{q,n\}^2))) +2n -(20(max\{q,n\}^2)) = q[12n+18+40(max\{q,n\}^2)]+2n-(20(max\{q,n\}^2))$. Then, in order for this solution to be optimal, one should have $q[12n+18+40(max\{q,n\}^2)]+2n-(20(max\{q,n\}^2)) \geq 40q(max\{q,n\}^2)+6qn+8q+n$. This can be reduced to $6qn+10q-20(max\{q,n\}^2) +n \geq 0$. But, one can easily check that for any $n\geq 3$ (which is always the case in 3SAT instances), this is not true; a contradiction. 
\end{proof}

\begin{lemma}\label{lem_size}
 Any optimal solution of $\lapcs$ problem on $(S_1,P_1)$ and $(S_2,P_2)$ is of length $40q(max\{q,n\}^2)+6qn+8q+n$.
\end{lemma}
\begin{proof}
By construction, in $S_1$ there is (1) $\forall 1\leq i\leq n$, $2q+1$ occurrences of $x_i$ (resp. $\overline{x_i}$); (2) $\forall 1\leq i\leq q$, $4$ occurrences of $Q_i$ (resp. $Q_{q+i}$); (3) $\forall 1\leq i\leq q$, $1$ occurrence of each $\{R_i^1,R_{q+i}^2,R_i^3,R_{q+i}^1,R_{q+i}^3,W_i,V_i,s_1,s_2,s_3\}$; (4) $\forall 1\leq i\leq q$, $2$ occurrences of $R_i^2$.

Whereas, in $S_2$, there is (1) $\forall 1\leq i\leq n$, $6q+1$ occurrences of $x_i$ (resp. $\overline{x_i}$); (2) $\forall 1\leq i\leq q$, $2$ occurrences of $Q_i$ (resp. $Q_{q+i}$); (3) $\forall 1\leq i\leq q$, $1$ occurrence of each $\{R_i^1,R_i^2,R_i^3,R_{q+i}^1,R_{q+i}^2,R_{q+i}^3,W_i,V_i,s_1,s_2,s_3\}$.

Therefore, in any optimal solution there may be only (1) $\forall 1\leq i\leq n$, $2q+1$ occurrences of $x_i$ (resp. $\overline{x_i}$); (2) $\forall 1\leq i\leq q$, $2$ occurrences of $Q_i$ (resp. $Q_{q+i}$); (3) $\forall 1\leq i\leq q$, $1$ occurrence of each $\{R_i^1,R_i^2,R_i^3,R_{q+i}^1,R_{q+i}^2,R_{q+i}^3,W_i,V_i,s_1,s_2,s_3\}$.

More precisely, by Lemma \ref{lem_canonicity}, and since, by construction, there is an arc in $P_2$ between $\chi(1,R_i^j,C^2_i)$ and $\chi(1,R_{q+i}^j,P^2_i)$, $\forall j \in\{1,2,3\}$, in any optimal solution, $\forall 1\leq i\leq q$, only half of the $\{R_i^1, R_i^2, R_i^3, R_{q+i}^1, R_{q+i}^2, R_{q+i}^3\}$ may be conserved. 

Moreover, any $x_i$ (resp. $\overline{x_i}$) of $S_1$ except in $C^1_q$, is linked by an arc to another $x_i$ (resp. $\overline{x_i}$), therefore by Lemma \ref{lem_canonicity}, in any optimal solution, $\forall 1\leq i\leq q-1$, only half of the occurrences of $x_i$ (resp. $\overline{x_i}$) may be conserved.

Finally, in any optimal solution, only half of the occurrences of $\{x_i,\overline{x_i}\}$ and one over $\{s_1,s_2,s_3\}$ in $C^1_q$ and $S_M^1$ may be conserved. Indeed, by construction, if this is not the case in $C^1_q$ (resp. $S_M^1$), it implies that at least one complete sequence $Q_q$ (resp. $V_1$ or $W_1$) is totally deleted -- which is not optimal since it is of length $n+1$ (resp. $20(max\{q,n\}^2)$).

On the whole, the maximal total length of any solution is thus equal to $40q(max\{q,n\}^2)+6qn+8q+n$. Moreover, this solution is composed of (1) $\forall 1\leq i\leq n$, $2q+1$ occurrences of either $x_i$ or $\overline{x_i}$, (2) $\forall 1\leq i\leq q$, $2$ occurrences of $Q_i$ and $Q_{q+i}$, (3) $\forall 1\leq i\leq q$, $1$ occurrence of each $\{W_i,V_i\}$ and either $s_1,$ $s_2$ or $s_3$ and (4) $\forall 1\leq i\leq q$, $R_i^{j_1},R_i^{j_2},R_{q+i}^{j_3}$ s.t. $\{j_1, j_2,j_3\}=\{1,2,3\}$.
\end{proof}

\begin{lemma}\label{lem_propag}
  In any optimal solution of $\lapcs$ problem on $(S_1,P_1)$ and $(S_2,P_2)$, if $\chi(1,x_k,S_M^1)$ (resp. $\chi(1,\overline{x_k},S_M^1)$) for a given $1\leq k\leq n$ is deleted then, $\forall 1\leq j\leq q$, $\chi(1,x_k,C_j^1)$ (resp. $\chi(1,\overline{x_k},C_j^1)$) is deleted.
\end{lemma}
\begin{proof}
By construction, $\forall 1\leq k \leq n$ only one of $\{x_k,\overline{x_k}\}$ may be conserved between $S_M^1$ and $S_M^2$ since $\chi(1,x_k,S_M^1) < \chi(1,\overline{x_k},S_M^1)$ whereas $\chi(1,\overline{x_k},S_M^2) < \chi(1,x_k,S_M^2)$. By Lemma \ref{lem_canonicity}, at least one symbol incident to any arc is deleted. Therefore, $\forall 1\leq k \leq n$ only one of $\{x_k,\overline{x_k}\}$ may be conserved between $C_1^1$ and $C_1^2$.

Let us suppose that for a given $1\leq k \leq n$, $\chi(1,\overline{x_k},S_M^1)$ is deleted. According to the proof of Lemma \ref{lem_size}, in any optimal solution, $\forall 1\leq k \leq n$ exactly one of $\{x_k,\overline{x_k}\}$ has to be deleted. Then $\chi(1,x_k,P_1^1)$ is deleted whereas $\chi(1,\overline{x_k},P_1^1)$ is conserved.

By construction, in $P^2_1$, since according to the proof of Lemma \ref{lem_size}, both occurrences of $Q_{q+1}$ and $R_1^{j_1},R_1^{j_2},R_{q+1}^{j_3}$ s.t. $\{j_1, j_2,j_3\}=\{1,2,3\}$ have to be conserved, either (1) $\{R_1^1,R_1^2,R_{q+1}^3\}$, (2) $\{R_1^1,R_1^3,R_{q+1}^2\}$ or (3) $\{R_1^2,R_1^3,R_{q+1}^1\}$ are conserved.

Let us first consider that $\{R_1^1,R_1^2,R_{q+1}^3\}$ are conserved. Then one can check that the only solution is to conserve $\chi(2,R_1^2,C_1^1)$ since otherwise at least half of the $x_k$'s would not be conserved. Consequently, the only solution is to conserve, $\forall 1\leq k \leq n$, the first (resp. last) occurrence of any $x_k$ or $\overline{x_k}$ in $C_1^2$ (resp. $P_1^2$) -- i.e. the occurrences appearing before $\chi(1,Q_1,C_1^2)$ (resp. after $\chi(2,Q_{q+1},P_1^2)$). Since by construction, there is an arc between $\chi(1,x_k,C_1^2)$ (resp. $\chi(1,\overline{x_k},C_1^2)$) and $\chi(3,x_k,P_1^2)$ (resp. $\chi(3,\overline{x_k},P_1^2)$), in order for $\chi(1,\overline{x_k},P_1^1)$ to be conserved, one has to conserved $\chi(3,\overline{x_k},P_1^2)$. Thus, by Lemma \ref{lem_canonicity}, $\chi(1,\overline{x_k},C_1^2)$ has to be deleted and, according to the proof of Lemma \ref{lem_size}, $\chi(1,x_k,C_1^2)$ has to be conserved.

Let us now consider that $\{R_1^1,R_1^3,R_{q+1}^2\}$ are conserved. By a similar reasoning, one can check that the only solution is to conserve, $\forall 1\leq k \leq n$, the second occurrence of any $x_k$ or $\overline{x_k}$ in $C_1^2$ (resp. $P_1^2$) -- i.e. the occurrences appearing between $\chi(1,Q_1,C_1^2)$ and $\chi(2,Q_1,C_1^2)$ (resp. $\chi(1,Q_{q+1},P_1^2)$ and $\chi(2,Q_{q+1},P_1^2)$). Since by construction, there is an arc between $\chi(2,x_k,C_1^2)$ (resp. $\chi(2,\overline{x_k},C_1^2)$) and $\chi(2,x_k,P_1^2)$ (resp. $\chi(2,\overline{x_k},P_1^2)$), in order to $\chi(1,\overline{x_k},P_1^1)$ to be conserved, one has to conserved $\chi(2,\overline{x_k},P_1^2)$. Thus, by Lemma \ref{lem_canonicity}, $\chi(2,\overline{x_k},C_1^2)$ has to be deleted and, according to the proof of Lemma \ref{lem_size}, $\chi(2,x_k,C_1^2)$ has to be conserved.

Finally, let us consider that $\{R_1^2,R_1^3,R_{q+1}^1\}$ are conserved. Once again, by a similar reasoning, one can check that the only solution is to conserve $\chi(1,R_1^2,C_1^1)$ since otherwise at least half of the $x_k$'s would not be conserved. Consequently, the only solution is to conserve, $\forall 1\leq k \leq n$, the last (resp. first) occurrence of any $x_k$ or $\overline{x_k}$ in $C_1^2$ (resp. $P_1^2$) -- i.e. the occurrences appearing after $\chi(2,Q_1,C_1^2)$ (resp. before $\chi(1,Q_{q+1},P_1^2)$). Since by construction, there is an arc between $\chi(3,x_k,C_1^2)$ (resp. $\chi(3,\overline{x_k},C_1^2)$) and $\chi(1,x_k,P_1^2)$ (resp. $\chi(1,\overline{x_k},P_1^2)$), in order to $\chi(1,\overline{x_k},P_1^1)$ to be conserved, one has to conserved $\chi(1,\overline{x_k},P_1^2)$. Thus, by Lemma \ref{lem_canonicity}, $\chi(3,\overline{x_k},C_1^2)$ has to be deleted and, according to the proof of Lemma \ref{lem_size}, $\chi(3,x_k,C_1^2)$ has to be conserved.

Therefore, in the three cases, if for a given $1\leq k \leq n$, $\chi(1,x_k,S_M^1)$ is conserved then so does $\chi(1,x_k,C_1^1)$. It is easy to see that, by a similar reasoning, if for a given $1\leq k \leq n$, $\chi(1,\overline{x_k},S_M^1)$ is conserved then so does $\chi(1,\overline{x_k},C_1^1)$.

With a similar reasoning, by reccurence, since, $\forall 1\leq i\leq q, 1\leq k\leq n$, there is an arc in $P_1$ between $\chi(1,x_k,C_i^1)$ (resp. $\chi(1,\overline{x_k},C_i^1)$) and $\chi(1,x_k,P_{i+1}^1)$ (resp. $\chi(1,\overline{x_k},P_{i+1}^1)$), if $\chi(1,x_k,C_i^1)$ is conserved then 
$\chi(1,x_k,P_{i+1}^1)$ is deleted. And therefore, with similar arguments, $\chi(1,x_k,C_{i+1}^1)$ is conserved. Once more, it is easy to see that this result still holds if $\chi(1,\overline{x_k},C_i^1)$ is conserved. 
\end{proof}

\begin{theorem}
 Given an instance of the problem 3SAT with $n$ variables and $q$ clauses, there exists a satisfying truth assignment iff the $\lapcs$ of $(S_1,P_1)$ and $(S_2,P_2)$ is of length $k' = 40q(max\{q,n\}^2)+6qn+8q+n$.
\end{theorem}
\begin{proof}
 ($\Rightarrow$) An optimal solution for $C_q=(x_1\vee x_2\vee \overline{x_3})\wedge(\overline{x_1}\vee \overline{x_2}\vee x_4)\wedge(x_2\vee \overline{x_3}\vee \overline{x_4})$ -- i.e. $x_1=x_3=true$ and $x_2=x_4=false$ -- is illustrated in Figures \ref{Fig_Example_1} and \ref{Fig_Example_2} where any symbol over a grey background have to be deleted. Suppose we have a solution of 3SAT, that is an assignment of each variable of $V_n$ satisfying $C_q$. Let us first list all the symbols to delete in $S_1$. 

For all $1\leq k \leq n$, if $x_k=false$ then delete, $\forall 1\leq j \leq q$,  $\{\chi(1,x_k,C^1_j),$ $\chi(1,\overline{x_k},P^1_j)\}$ and $\chi(1,x_k,S^1_M)$; otherwise delete, $\forall 1\leq j \leq q$,  $\{\chi(1,\overline{x_k},C^1_j),$ $\chi(1,x_k,P^1_j)\}$ and $\chi(1,\overline{x_k},S^1_M)$.

For each $L_i^j$ satisfying $c_i$ with the biggest index $j$ with $1\leq i \leq q$, 

if (1) $j=1$ then delete $\{\chi(1,R^3_i,C^1_i),$ $\chi(1,Q_i,C^1_i),$ $\chi(1,R^2_i,C^1_i),$ $\chi(2,Q_i,C^1_i),$ $\chi(1,s_2,C^1_i),$ $\chi(1,s_3,C^1_i),$ $\chi(1,R^2_{q+i},P^1_i),$ $\chi(1,R^1_{q+i},P^1_i),$ $\chi(3,Q_{q+i},P^1_i),$ \\$\chi(4,Q_{q+i},P^1_i)\}$ (cf Figure \ref{Fig_Example_1}.a); 

if (2) $j=2$ then delete $\{\chi(1,R^2_i,C^1_i),$ $\chi(2,Q_i,C^1_i),$  $\chi(1,s_1,C^1_i),$ $\chi(1,s_3,C^1_i),$ $\chi(3,Q_i,C^1_i),$ $\chi(2,R^2_i,C^1_i),$ $\chi(2,Q_{q+i},P^1_i),$ $\chi(1,R^3_{q+i},P^1_i),$  $\chi(1,R^1_{q+i},P^1_i),$ \\$\chi(3,Q_{q+i},P^1_i)\}$ (cf Figure \ref{Fig_Example_2}.a); 

if (3) $j=3$ then delete $\{\chi(1,s_1,C^1_i),$ $\chi(1,s_2,C^1_i),$ $\chi(3,Q_i,C^1_i),$ $\chi(2,R^2_i,C^1_i),$ $\chi(4,Q_i,C^1_i),$ $\chi(1,R^1_i,C^1_i),$ $\chi(1,Q_{q+i},P^1_i),$ $\chi(2,Q_{q+i},P^1_i),$  $\chi(1,R^3_{q+i},P^1_i),$ \\$\chi(1,R^2_{q+i},P^1_i)\}$ (cf Figure \ref{Fig_Example_2}.c);

Let us now list all the symbols in $S_2$ to be deleted. 

For all $1\leq k \leq n$, if $x_k=false$ then delete $\chi(1,x_k,S^2_M)$; otherwise delete $\chi(1,\overline{x_k},S^2_M)$.

For each $L_i^j$ satisfying $c_i$ with the biggest index $j$ with $1\leq i \leq q$, 

if (1) $j=1$ then delete $\forall 1\leq k \leq n$ $\{\chi(1,R^3_i,C^2_i),$ $\chi(1,s_2,C^2_i),$ $\chi(2,x_k,C^2_i),$ $\chi(2,\overline{x_k},C^2_i),$ $\chi(1,s_3,C^2_i),$ $\chi(3,x_k,C^2_i),$ $\chi(3,\overline{x_k},C^2_i),$ $\chi(1,x_k,P^2_i),$ $\chi(1,\overline{x_k},P^2_i),$ $\chi(1,R^1_{q+i},P^2_i),$ $\chi(1,R^2_{q+i},P^2_i),$ $\chi(2,x_k,P^2_i),$ $\chi(2,\overline{x_k},P^2_i)\}$. Moreover, if $x_k=false$ with $1\leq k \leq n$ then delete, $\{\chi(1,x_k,C^2_i),$ $\chi(3,\overline{x_k},P^2_i)\}$; otherwise delete \\$\{\chi(1,\overline{x_k},C^2_i),$ $\chi(3,x_k,P^2_i)\}$ (cf Figure \ref{Fig_Example_1}.a); 

if (2) $j=2$ then delete $\forall 1\leq k \leq n$ $\{\chi(1,R^2_i,C^2_i),$ $\chi(1,s_1,C^2_i),$ $\chi(1,x_k,C^2_i),$ $\chi(1,\overline{x_k},C^2_i),$ $\chi(1,s_3,C^2_i),$ $\chi(3,x_k,C^2_i),$ $\chi(3,\overline{x_k},C^2_i),$ $\chi(1,x_k,P^2_i),$ $\chi(1,\overline{x_k},P^2_i),$ $\chi(1,R^1_{q+i},P^2_i),$ $\chi(1,R^3_{q+i},P^2_i),$ $\chi(3,x_k,P^2_i),$ $\chi(3,\overline{x_k},P^2_i)\}$. Moreover, if $x_k=false$ with $1\leq k \leq n$ then delete, $\{\chi(2,x_k,C^2_i),$ $\chi(2,\overline{x_k},P^2_i)\}$; otherwise delete \\$\{\chi(2,\overline{x_k},C^2_i),$ $\chi(2,x_k,P^2_i)\}$ (cf Figure \ref{Fig_Example_2}.a); 

if (3) $j=3$ then delete $\forall 1\leq k \leq n$ $\{\chi(1,R^1_i,C^2_i),$ $\chi(1,s_1,C^2_i),$ $\chi(1,x_k,C^2_i),$ $\chi(1,\overline{x_k},C^2_i),$ $\chi(1,s_2,C^2_i),$ $\chi(2,x_k,C^2_i),$ $\chi(2,\overline{x_k},C^2_i),$ $\chi(2,x_k,P^2_i),$ $\chi(2,\overline{x_k},P^2_i),$ $\chi(1,R^2_{q+i},P^2_i),$ $\chi(1,R^3_{q+i},P^2_i),$ $\chi(3,x_k,P^2_i),$ $\chi(3,\overline{x_k},P^2_i)\}$. Moreover, if $x_k=false$ with $1\leq k \leq n$ then delete, $\{\chi(3,x_k,C^2_i),$ $\chi(1,\overline{x_k},P^2_i)\}$; otherwise delete \\$\{\chi(3,\overline{x_k},C^2_i),$ $\chi(1,x_k,P^2_i)\}$ (cf Figure \ref{Fig_Example_2}.c);

By construction, the natural order of the symbols of $S_1$ and $S_2$ allows the corresponding set of undeleted symbols to be conserved in a common arc-preserving common subsequence between $(S_1,P_1)$ and $(S_2,P_2)$. Let us now prove that the length of this last is $k'$. One can easily check that this solution is composed of $\forall 1\leq k\leq n$, (1) $2q+1$ occurrences of either $x_k$ or $\overline{x_k}$, (2) $\forall 1\leq i\leq q$, $2$ occurrences of $Q_i$ and $Q_{q+i}$, (3) $\forall 1\leq i\leq q$, $1$ occurrence of each $\{W_i,V_i\}$ and either $s_1,$ $s_2$ or $s_3$ and (4) $\forall 1\leq i\leq q$, $R_i^{j_1},R_i^{j_2},R_{q+i}^{j_3}$ s.t. $\{j_1, j_2,j_3\}=\{1,2,3\}$. Thus, the length of the solution is $40q(max\{q,n\}^2)+6qn+8q+n$.

($\Leftarrow$) Suppose we have an optimal solution -- i.e. a set of symbols $S_d$ to delete -- for $\lapcs$ of $(S_1,P_1)$ and $(S_2,P_2)$. Let us define the truth assignment of $V_n$ s.t., $\forall 1\leq i \leq q$, if $\chi(1,s_j,C^1_i)\not\in S_d$ then $L_i^j$ is true. Let us prove that it is a solution of 3SAT.

By construction, if $L_i^j=x_k$ (resp. $\overline{x_k}$) then in $C^1_i$, $s_j$ appears between $x_k$ and $\overline{x_k}$ whereas in $C^2_j$ it appears after $\overline{x_k}$ (resp. before $x_k$). Thus, if $\chi(1,s_j,C^1_i)$ is not deleted then $\overline{x_k}$ (resp. $x_k$) in $C^1_i$ is deleted if $L_i^j=x_k$ (resp. $\overline{x_k}$). Consequently, according to the proof of Lemma \ref{lem_propag}, if $\chi(1,s_j,C^1_i)$ is not deleted then $\overline{x_k}$ (resp. $x_k$) in all $C^1_{i'}$, with $1\leq i'\leq q$ is deleted if $L_i^j=x_k$ (resp. $\overline{x_k}$). Therefore, we can ensure that one cannot obtain $L_i^j$ and $L_{i'}^{j'}$ being true whereas $L_i^j=\overline{L_{i'}^{j'}}$ (that is a variable cannot be simultaneously true and false). By Lemma \ref{lem_size}, we can ensure that for any $1\leq i \leq q$ exactly one of $\{s_1,s_2,s_3\}$ is conserved in $C^1_i$. Therefore, for any clause $c_i$ at least one of its literal is set to true. This ensures that our solution is a solution of 3SAT.
\end{proof}

\section{Future work}

From a computational biology point of view, especially for comparing stems, one may, however, be mostly interested in the case $k$ (length of the common subsequence searched) might not be
assumed to small compared to $n$.
A first approach is provided in
\cite{Alber:Gramm:Guo:Niedermeier:TCS:2004}
where it is proved that,
given two sequences of length at most $n$ and nested
arc structure, an arc-preserving common subsequence can be determined (if it exists) 
in $O(3.31^{k_1 + k_2} \, n)$ time; obtained
by deleting (together with corresponding arcs) $k_1$ letters from the
first and $k_2$ letters from the second sequence.
Improving the running time of the parameterization in case of stem arc
structures appears to be a promising line of research.

\bibliographystyle{agsm}
\bibliography{biblio}

\end{document}